\documentclass[a4paper]{article}
\usepackage{authblk}
\usepackage{amsmath}
\usepackage{amsthm}
\usepackage{booktabs}
\usepackage{url}

\usepackage{amssymb}

\usepackage[figuresright]{rotating}
\usepackage{tcolorbox}

\usepackage[font=footnotesize,labelfont=bf]{caption}
\usepackage[font=footnotesize,labelfont=bf]{subcaption}

\usepackage[english]{babel}
\begin{document}
\title{An operational point of view to the theory of multi-variable/multi-index Hermite polynomials}
\author[1]{Giuseppe Dattoli}
\author[2]{Silvia Licciardi}
\author[1]{Elio Sabia\thanks{corresponding author}}
\affil[1]{\textit{ENEA---Frascati Research Center, Via Enrico Fermi 45, 00044 Rome, Italy; pinodattoli@libero.it; elio.sabia@gmail.com}  }
\affil[2]{\textit{Università degli Studi di Palermo, Dipartimento di Ingegneria, Viale delle Scienze, 90128, Palermo, Italy; silvia.licciardi@unipa.it}}
\maketitle


\abstract{The use of algebraic tools of operational and umbral nature is exploited to develop  a new point of view and to extend the theory of Hermite polynomials, with more than one variable also of complex nature. The techniques we adopt includes multivariable/many index Hermite- Kampè-dè-Fèrièt polynomials of order two and higher. It will be shown that the treatment, foreseen here, simplifies the study of the relevant properties and the associated computational technicalities.}


\section{Introduction}

Two variable Hermite – Kampè –dè Fèrièt (HKdF) polynomials of order 2 \cite {App-de-Fer}

\begin{equation}
\label{2_var_HKdf_2_ord}
H_{n}(x,y)=n!\sum_{r=0}^{\lfloor \frac{n}{2}\rfloor}\frac{x^{n-2r}y^r}{(n-2r)!r!}
\end{equation}

\noindent with generating function 

\begin{equation}
\label{2_var_HKdf_2_ord_gen_func}
\sum_{n=0}^{\infty}\frac{t^n}{n!}H_{n}(x,y)=e^{xt+yt^2}
\end{equation}

\noindent have been shown to be characterized by a wealth of properties \cite{App-de-Fer}, \cite{bab-dat-lic-sab}, which have provided a significantly simplifying tool to study their properties, get a more thorough understanding of the underlying theory and enter deeply into the their link with other families of special polynomials and functions \cite{dat}. The ordinary counterparts are just a particular case of the HKdF and indeed we have \cite{Andrews}

\begin{equation}\label{part_case_HKdf}
\begin{split}
& H_{n}(x,-\frac{1}{2})= He_n(x)\\
& H_{n}(2x,-1)=H_{n}(x)
\end{split}
\end{equation}

\noindent  From eq. \eqref{2_var_HKdf_2_ord} we find that, by keeping $y=0$, the polynomials reduce the corresponding monomial in $x$, namely

\begin{equation}\label{monom}
H_{n}(x,0)=x^n
\end{equation}
                                      
\noindent More in general it has also been shown that they are quasi-monomials \cite{dat_1} according to the relationships

\begin{equation}\label{deriv_HKdf}
\begin{split}
& \hat{P}H_{n}(x,y)=n H_{n-1}(x,y)\\
& \hat{M}H_{n}(x,y)= H_{n+1}(x,y)
\end{split}
\end{equation}
                    
\noindent Where $\hat{P}$ ,  $\hat{M}$  are derivative and multiplicative operators defined as

\begin{equation}\label{oper_der_mult}
\begin{split}
& \hat{P}= \partial_x\\
& \hat{M}= x+2y\partial_x
\end{split}
\end{equation}

\noindent and eventually yields the relevant differential equations satisfied by this family of polynomials

\begin{equation}\label{op_diff_eq_poly}
\hat{M}\hat{P}H_{n}(x,y)=nH_{n}(x,y)
\end{equation}

\noindent which, in differential form, reads

\begin{equation}\label{diff_eq_poly}
x\partial_xH_{n}(x,y) + 2y\partial_{x}^{2}H_{n}(x,y)=nH_{n}(x,y)
\end{equation}

\noindent If $y$ is treated as an ordinary constant (keeping e. g. $y=-1/2$) eq. \eqref{diff_eq_poly} reduces to the second order ODE

\begin{equation}\label{ord_diff_eq}
\begin{split}
& z''-xz'+nz= 0\\
& z(x)= He_n(x)
\end{split}
\end{equation}

\noindent The HKdF polynomials have also appeared in the mathematical literature under the name of Heat polynomials \cite{Wid}, since they are a “natural” solution of the Heat equation

\begin{equation}\label{Heat_eq}
\begin{split}
& \partial_yH_n(x,y)= \partial_{x}^{2}H_n(x,y)\\
& H_n(x,0)= x^n
\end{split}
\end{equation}

\noindent The previous equation is a straightforward Cauchy problem, which yields, as corollary, the following operational definition \cite{bab-dat-lic-sab}

\begin{equation}\label{Op_def_Heat_eq}
H_n(x,y)=e^{y\partial_{x}^{2}}x^n
\end{equation}

\noindent The higher order Hermite \cite{bab-dat-lic-sab}

\begin{equation}\label{high_ord_Herm}
H_{n}^{(m)}(x,y)=n!\sum_{r=0}^{\lfloor \frac{n}{m}\rfloor}\frac{x^{n-mr}y^r}{(n-mr)!r!}
\end{equation}

\noindent with generating function

\begin{equation}\label{gen_func_high_ord_Herm}
\sum_{n=0}^{\infty}\frac{t^n}{n!}H_{n}^{(m)}(x,y)=e^{xt+yt^m}
\end{equation}

\noindent satisfy the generalized heat equation \cite{bab-dat-lic-sab},\cite{Wid} 

\begin{equation}\label{gen_Heat_eq}
\begin{split}
& \partial_yH_{n}^{(m)}(x,y)= \partial_{x}^{m}H_{n}^{(m)}(x,y)\\
& H_{n}^{(m)}(x,0)= x^n
\end{split}
\end{equation}

\noindent and the operational definition

\begin{equation}\label{op_gen_Heat_eq}
H_{n}^{(m)}(x,y)=e^{y\partial_{x}^{m}}x^n
\end{equation}

\noindent In this article, we discuss the extension of the previous properties to the case of Hermite polynomials with complex variables. We start by considering polynomials of the type

\begin{equation}\label{complex_Herm_pol}
\begin{split}
& H_{n}(z,y) \to z= x_1+ix_2 \in C, y\in R\\
& H_{n}(z,w) \to z=x_1+ix_2 \in C, w=y_1+iy_2 \in C
\end{split}
\end{equation}

\noindent namely, order 2 HKdF with one complex and one real and with to complex variables.
The first family (with $y$ a negative constant) has been used to present a deeper formulations of the theory of coherent states\cite{Gaz}\cite{Nic}\cite{Ism}.
It is evident that, a natural extension of the operational definition in eq. \eqref{Op_def_Heat_eq}, yields

\begin{equation}\label{extens_op_def}
H_{n}(z,y)=e^{ix_2\partial_{x_1}}H_n(x_1,y)=e^{ix_2\partial_{x_1}+y\partial_{x_1}^2}x_1^n
\end{equation}

\noindent which can be exploited to derive the relevant generating function (namely the same in eq.\eqref{2_var_HKdf_2_ord} with $z$ instead of $x$) and the alternative definition in terms of series \footnote{Note that $e^{ix_2\partial_{x_1}+y\partial_{x_1}^2}=\sum_{r=0}^{\infty}\frac{H_r(ix_2,y)}{r!}\partial_{x_1}^r$}

\begin{equation}\label{complex_ser_def}
H_{n}(z,y)=n!\sum_{r=0}^{n}\frac{H_r(ix_2,y)}{r!(n-r)!}x_1^{n-r}
\end{equation}
 
\noindent Before going further into specific details, we provide an idea of the formalism we are going to apply, by studying examples of how the previous notions are helpful in computations.  We consider therefore the following one dimensional improper integral

\begin{equation}\label{Imp_Integ_Herm_pol}
\begin{split}
& I_{n}(x,b) =  \int_{-\infty}^{+\infty}H_n(x,y)e^{-by^2}dy\\
& Re(b)>0
\end{split}
\end{equation}

\noindent and use  the operational definition in eq.\eqref{Op_def_Heat_eq} to rewrite the previous integral as

\begin{equation}\label{Op_def_Imp_Integ_Herm_pol}
\begin{split}
& I_{n}(x,b) = \hat{E}(y,b)x^n\\
& \hat{E}(y,b) =  \int_{-\infty}^{+\infty}e^{y\partial_x^2}e^{-by^2}dy\\
\end{split}
\end{equation}

\noindent The operator $\hat{E}(y,b)$ when acting on the monomial $x^n$ , defines a further family of Hermite and can be worked out in a non-integral form, by treating it as a conventional Gaussian integration, namely by keeping the second derivative in the exponential as an ordinary constant and obtaining

\begin{equation}\label{new_op}
\hat{E}(y,b)= \sqrt{\frac{\pi}{b}}e^{\frac{1}{4b}\partial_x^4}
\end{equation}

\noindent Therefore, on account of the previous \eqref{op_gen_Heat_eq}, we eventually find

\begin{equation}\label{Int_eval_new_op}
I_n(x,b)= \sqrt{\frac{\pi}{b}}e^{\frac{1}{4b}\partial_x^4}x^n=\sqrt{\frac{\pi}{b}}H_n^{(4)}\left(x,\frac{1}{4b}\right)
\end{equation}

\noindent The conclusion we have just obtained can be stated with other means, but the procedure, we have followed is remarkable, in its simplicity.\\
The second example involves the evaluation of integrals of a complex two variable Hermite, namely

\begin{equation}\label{Int_imp_cpx_Herm}
I_n(\delta_1,\delta_2,a,b,y)= \int_{-\infty}^{+\infty}H_n(x_1+\delta_1+i(x_2+\delta_2),y)e^{-ax_1^2-bx_2^2}dx_1dx_2
\end{equation}

\noindent (note that we have indicated the double integral with $\int_{-\infty}^{+\infty}F(x_1,x_2)e^{-ax_1^2-bx_2^2}dx_1dx_2$ instead of using  $\int_{R^2}F(x_1,x_2)e^{-ax_1^2-bx_2^2}dx_1dx_2$)\\
\noindent which can be worked out by the use of the generating function method, developed in \cite{bab-dat-lic-sab}. Multiplying indeed both sides of \eqref{Int_imp_cpx_Herm} by $t^n$ and summing up over the index $n$, we find

\begin{equation}\label{Int_imp_cpx_Herm_mult}
\begin{split}
& \sum_{n=0}^{\infty}\frac{t^n}{n!}I_n(\delta_1,\delta_2,a,b,y)= e^{(\delta_1+i\delta_2)t+yt^2} \int_{-\infty}^{+\infty}e^{(x_1+ix_2)t}e^{-ax_1^2-bx_2^2}dx_1dx_2=\\
& =\frac{\pi}{\sqrt{ab}}e^{(\delta_1+i\delta_2)t+\left(y+\frac{1}{4a}-\frac{1}{4b}\right)t^2}
\end{split}
\end{equation}

\noindent The exponential function on the rhs of the previous identity can be expanded in HKdF of order 2, namely

\begin{equation}\label{Expansion_HKdf_ord_2}
e^{(\delta_1+i\delta_2)t+\left(y+\frac{1}{4a}-\frac{1}{4b}\right)t^2}=\sum_{n=0}^{\infty}\frac{t^n}{n!}H_n\left(\delta_1+i\delta_2,y+\frac{1}{4a}-\frac{1}{4b} \right)
\end{equation}

\noindent and after equating the $t$-like powers in eq. \eqref{Expansion_HKdf_ord_2} we can conclude that

\begin{equation}\label{Int_imp_cpx_Herm_solut}
I_n(\delta_1,\delta_2,a,b,y)= \frac{\pi}{\sqrt{ab}}H_n\left(\delta_1+i\delta_2,y+\frac{1}{4a}-\frac{1}{4b}\right)
\end{equation}

\noindent In order to underscore the implications offered by the formalism, we discuss the extension of the previous example to the case (same notation as before for the integral in $R^3$ )

\begin{equation}\label{Int_imp_cpx_Herm_in_R3}
I_n(\delta_1,\delta_2,a,b,c)= \int_{-\infty}^{+\infty}H_n(x_1+\delta_1+i(x_2+\delta_2),y)e^{-ax_1^2-bx_2^2-cy^2}dx_1dx_2dy
\end{equation}

(note that  $\int_{R^3}F(x_1,x_2,x_3)e^{-ax_1^2-bx_2^2-cx_3^2}dx_1dx_2dx_3 \to $\\
\hspace*{1.8cm}$\to \int_{-\infty}^{+\infty}F(x_1,x_2,x_3)e^{-ax_1^2-bx_2^2-cx_3^2}dx_1dx_2dx_3$  )\\
\noindent can be evaluated by the use of the same procedure, which eventually yields

\begin{equation}\label{Int_imp_cpx_Herm_mult_1}
\begin{split}
& \sum_{n=0}^{\infty}\frac{t^n}{n!}I_n(\delta_1,\delta_2,a,b,c)= e^{(\delta_1+i\delta_2)t} \int_{-\infty}^{+\infty}e^{(x_1+ix_2)t+yt^2}e^{-ax_1^2-bx_2^2-cy^2}dx_1dx_2dy=\\
& =\frac{\pi^{3/2}}{\sqrt{abc}}e^{(\delta_1+i\delta_2)t+\frac{1}{4}\left(\frac{1}{a}-\frac{1}{b}\right)t^2+\frac{1}{4c}t^4}
\end{split}
\end{equation}

\noindent The integral can be written in a closed form, using a slightly more general form of the previously introduced higher order Hermite. The exponential function on the rhs of eq. \eqref{Int_imp_cpx_Herm_mult_1} is the generating function of HKdF polynomials of order 4. They are multivariable HKdF, defined by the generating function \cite{dat_2}

\begin{equation}\label{Higher_ord_Herm_gen_fun}
\sum_{n=0}^{\infty}\frac{t^n}{n!}H_n^{(4)}(x_1,x_2,x_3,x_4) = e^{\sum_{s=1}^{4}x_st^s}
\end{equation}

\noindent  where

\begin{equation}\label{Higher_ord_Herm}
\begin{split}
& H_n^{(4)}(x_1,x_2,x_3,x_4)= n!\sum_{r=0}^{\lfloor \frac{n}{4}\rfloor}\frac{H_{n-4r}^{(3)}(x_1,x_2,x_3)x_4^r}{(n-4r)!r!}\\
& H_n^{(3)}(x_1,x_2,x_3)= n!\sum_{r=0}^{\lfloor \frac{n}{3}\rfloor}\frac{H_{n-3r}(x_1,x_2)x_3^r}{(n-3r)!r!}
\end{split}
\end{equation}

\noindent They satisfy the obvious (but important property)\footnote{The superscript $(2)$ is not explicitly mentioned, unless specifically needed.}

\begin{equation}\label{Higher_ord_Herm_prop}
\begin{split}
& H_n^{(4)}(x_1,x_2,x_3,0)= H_n^{(3)}(x_1,x_2,x_3)\\
& H_n^{(4)}(x_1,x_2,0,0)=H_n(x_1,x_2)
\end{split}
\end{equation}

\noindent According to eqs. \eqref{Higher_ord_Herm_gen_fun} and \eqref{Int_imp_cpx_Herm_mult_1}, the integral \eqref{Int_imp_cpx_Herm_in_R3} can eventually be written as
\begin{equation}\label{Int_as_H4}
I_n(\delta_1,\delta_2,a,b,y)=\frac{\pi^{3/2}}{\sqrt{abc}}H_n^{(4)}\left(\delta_1+i\delta_2, \frac{1}{4a}-\frac{1}{4b},0,\frac{1}{4c}\right) 
\end{equation}

\noindent Let us consider the case of two complex variables and note that 

\begin{equation}\label{2_complex_var_Herm}
\begin{split}
& \sum_{n=0}^{\infty}\frac{t^n}{n!}H_n(z,w)=e^{zt+wt^2}= e^{x_1t+y_1t^2}e^{ix_2t+iy_2t^2}=\sum_{s=0}^{n} H_n(x_1,y_1| ix_2,iy_2)\\
& H_n(x_1,y_1| ix_2,iy_2)=\sum_{s=0}^{n}\binom{n}{s}H_{n-s}(x_1,y_1)H_s(ix_2,iy_2)=\\
& =e^{-iy_2\partial_{x_2}^2+y_1\partial_{x_1}^2}(x_1+ix_2)^n
\end{split}
\end{equation}

\noindent and the use of the integration procedure, developed before, leads to

\begin{equation}\label{Int_as_H4_two_compl_var}
I_n(\delta_1,\delta_2,a,b,c,d)=\frac{\pi^2}{\sqrt{abcd}}H_n^{(4)}\left(\delta_1+i\delta_2, \frac{1}{4}\frac{b-a}{ab},0,\frac{1}{4}\frac{d-c}{cd}\right) 
\end{equation}

\noindent In this introductory section we have fixed the main lines of the formalism, we will develop in the forthcoming sections, where it will be exploited to treat , among the other things, integrals involving products of two complex variables Hermite polynomials.

\section{Two-index Complex Hermite}
Before entering into the specific details of this section, we introduce the two index/two variable Hermite, originally proposed by Hermite himself \cite{Hermite}\cite{App-de-Fer} and reformulated, in a different context, in \cite{Dat-Tor}\cite{bab-dat-lic-sab} where they have been defined as

\begin{equation}\label{two_ind_cpl_her}
\begin{split}
&H_{m,n}(x,y;z,w | \tau)=m!n!\sum_{r=0}^{min[m,n]}\frac{\tau^rH_{m-r}(x,y)H_{n-r}(z,w)}{r!(m-r)!(n-r)!}\\
& x,y,z,w,\tau \in C
\end{split}
\end{equation}

\noindent The relevant generating function is provided by the following extension of the single index counterpart

\begin{equation}\label{Gen_fun_Hn_two_compl_var}
\sum_{m,n=0}^{\infty}\frac{u^m}{m!}\frac{v^n}{n!}H_{m,n}(x,y;z,w | \tau) = e^{xu+yu^2+zv+wv^2+\tau uv} 
\end{equation}

\noindent where “tau” has the role of entanglement between the couple of variables $(x, y)$ and $(z, w)$.  In operational form the polynomials in eq. \eqref{two_ind_cpl_her} can be written as

\begin{equation}\label{two_ind_cpl_her_op_form}
\begin{split}
& H_{m,n}(x,y;z,w | \tau)=e^{y\partial_x^2+w\partial_y^2}h_{m,n}(x,z|\tau)\\
& h_{m,n}(x,z | \tau)=m!n!\sum_{r=0}^{min[m,n]}\frac{\tau^rx^{m-r}z^{n-r}}{r!(m-r)!(n-r)!}\\
& h_{m,n}(x,z|\tau)=e^{\tau\partial_{x,y}}(x^my^n)
\end{split}
\end{equation}

\noindent where $h_{m,n}(x,z|\tau)$  are the incomplete Hermite polynomials discussed in \cite{Ito}\cite{Wun}.\\
These elementary properties are useful for practical purposes. Let us now consider the integral

\begin{equation}\label{Int_two_ind_cpl_her}
\begin{split}
& I_{m,n}(\delta,\eta,y,w,a)=\int_{-\infty}^{+\infty]}H_{m}(x+\delta,y)H_{n}(x+\eta,w)e^{-ax^2}dx\\
& Re(a)>0,\; \delta,\eta, y,w \in C
\end{split}
\end{equation}

\noindent and note that it can be  explicitly worked out by the use of the previously foreseen procedure, employing the generating function method. We find indeed

\begin{equation}\label{Gen_fun_two_ind_cpl_her}
\begin{split}
& \sum_{m,n=0}^{\infty}\frac{u^m}{m!}\frac{v^n}{n!}I_{m,n}(\delta,\eta,y,w,a)=e^{yu^2+wv^2+\delta u+\eta v}\int_{-\infty}^{+\infty}e^{-ax^2+x(u+v)}dx=\\
& \sqrt{\frac{\pi}{a}}e^{\delta u+\eta v+\left(y+\frac{1}{4a}\right)u^2+\left(w+\frac{1}{4a} \right)v^2+\frac{uv}{2a}}
\end{split}
\end{equation}

\noindent And eventually 

\begin{equation}\label{Int_Hn_two_compl_var}
I_{m,n}(\delta,\eta,y,w,a)=\sqrt{\frac{\pi}{a}}H_{m,n}\left(\delta,y+\frac{1}{4a};\eta, w+\frac{1}{4a}| \frac{1}{2a}\right)
\end{equation}

\noindent The extension of the previous result to the complex Hermite case is simply given by 

\begin{equation}\label{Int_two_ind_cpl_her_41}
\begin{split}
& I_{m,n}(\delta_1,\delta_2,y_1,y_2,a,b)=\\
& =\int_{-\infty}^{+\infty}H_m(x_1+\delta_1+i(x_2+\delta_2),y_1)H_n(x_1+\delta_1+i(x_2+\delta_2),y_2)e^{-ax_1^2-bx_2^2}dx_1dx_2\\
& Re(a), \; Re(b)>0
\end{split}
\end{equation}

\noindent Which in terms of multi-index Hermite reads

\begin{equation}\label{Int_two_ind_cpl_her_42}
\begin{split}
& I_{m,n}(\delta_1,\delta_2,y_1,y_2,a,b)=\\
& =\frac{\pi}{\sqrt{ab}}H_{m.n}\left(\delta_1+i\delta_2,y_1+\frac{1}{4}\frac{b-a}{ab},\delta_1+i\delta_2,y_2+\frac{1}{4}\frac{b-a}{ab}|\frac{1}{2}\frac{b-a}{ab}\right)
\end{split}
\end{equation}
 
\noindent This is a general result,  which follows from the request of convergence of the integral, ensured by the conditions on the constants $a, b$.\\
In the case in which the Hermite’s appearing in the definition of the integral in eq.\eqref{Int_as_H4_two_compl_var} are the ordinary family we have the correspondence

\begin{equation}\label{ord_fam_Her}
H_n(2x,-1)=2^nH_n\left(x,-\frac{1}{4}\right)
\end{equation}
 
\noindent And, if the following conditions are satisfied 

\begin{equation}\label{Her_44}
Re(a),\; Re(b)>0, \; 0<a<b; \; \frac{1}{a}=1+\frac{1}{b},\; \delta_{1,2}=0
\end{equation}
  
\noindent we find, on account of eqs. \eqref{Gen_fun_two_ind_cpl_her}\eqref{Int_two_ind_cpl_her_41} the “orthogonality” identity

\begin{equation}\label{Her_45}
\begin{split}
& I_{m,n}\left(0,0,-\frac{1}{4},-\frac{1}{4},a,a\right)=\pi\frac{2^{-n}n!}{a\sqrt{\frac{1}{1-a}}}\delta_{m,n}\\
& \delta_{m,n} \; \equiv \; \mbox{Kronecker delta} 
\end{split}
\end{equation}
 
\noindent a result already well-established in the literature (see \cite{Ism} and references therein).\\
We can now play with the operational rules, we established before, to work out Gaussian integrals involving two index Hermite. Regarding e. g. the two index case, the use of operational rule \eqref{two_ind_cpl_her_op_form}, allows to transform the integral

\begin{equation}\label{Int_46}
I_{m,n}(x,z;a,b|\tau)=\int_{-\infty}^{+\infty}H_{m,n}(x,y;z,w|\tau)e^{-ay^2-bw^2}dydw
\end{equation}
 
\noindent into

\begin{equation}\label{Int_47}
I_{m,n}(x,z;a,b|\tau)=\int_{-\infty}^{+\infty}e^{y\partial_x^2+w\partial_z^2}e^{-ay^2-bw^2}dydw \; h_{m,n}(x,z | \tau)
\end{equation}
 
\noindent The use of the previously outlined integration procedure, yields

\begin{equation}\label{Int_48}
I_{m,n}(x,z;a,b |\tau)=\frac{\pi}{\sqrt{ab}}e^{\frac{1}{4}\left(\frac{1}{a}\partial_x^4
+\frac{1}{b}\partial_z^4\right)} h_{m,n}(x,z | \tau)=\frac{\pi}{\sqrt{ab}}H_{m,n}^{(4,4,1)}\left( x,\frac{1}{4a};z,\frac{1}{4b}|\tau \right)
\end{equation}
 
\noindent The explicit form of the (4,4,1) polynomials of order $m,n$ is obtained quite straightforwardly by the use of the operational identity \eqref{op_gen_Heat_eq}, namely

\begin{equation}\label{Her_pol_ord_441}
H_{m,n}^{(4,4,1)}(x_1,x_2;x_3,x_4 | \tau)=m!n!\sum_{r=0}^{min[m,n]}\frac{\tau^r H_{m-r}^{(4)}(x_1,x_2)H_{n-r}^{(4)}(x_3,x_4)}{r!(m-r)!(n-r)!}
\end{equation}

\noindent A further use of the previous formalism allows the derivation of the integral reported below including Gaussian integration on complex variables. Namely

\begin{equation}\label{Her_pol_eq_50}
\begin{split}
& \int_{-\infty}^{+\infty}H_m(x_1,y_1+iy_2)H_n(x_2,y_1-iy_2)e^{-ay_1^2-by_2^2}dy_1dy_2=\\
& =\int_{-\infty}^{+\infty}e^{(y_1+iy_2)\partial_{x_1}^2+(y_1-iy_2)\partial_{x_1}^2-ay_1^2-by_2^2}dy_1dy_2(x_1^mx_2^n)
\end{split}
\end{equation}

\noindent By applying the same procedure as before, namely performing the Gaussian integrals by treating second derivatives as ordinary algebraic quantities, we end up with

\begin{equation}\label{Her_pol_eq_51}
\begin{split}
& \int_{-\infty}^{+\infty}H_m(x_1,y_1+iy_2)H_n(x_2,y_1-iy_2)e^{-ay_1^2-by_2^2}dy_1dy_2=\\
& =\frac{\pi}{\sqrt{ab}}H_{m,n}^{(4,4,2)}\left(x_1,\frac{1}{2}\sigma_-;x_2,\frac{1}{2}\sigma_{-} | \sigma_{+} \right)\\
& \sigma_{\pm}=\frac{1}{2}\left(\frac{1}{a}\pm \frac{1}{b} \right)\\
& H_{m,n}^{(4,4,2)}(x_1,x_2;x_3,x_4 | \tau)=m!n!\sum_{r=0}^{min[\lfloor \frac{m}{2} \rfloor,\lfloor \frac{n}{2} \rfloor]} \frac{\tau^rH_{m-2r}^{(4)}(x_1,x_2)H_{n-2r}^{(4)}(x_3,x_4)}{r!(m-2r)!(n-2r)!}
\end{split}
\end{equation}
  
\noindent In this section we have accomplished some steps towards the integration of Gaussian integrals containing different forms of Hermite like polynomials, with real and complex variable. In  the forthcoming sections we move to more advanced computations.

\section{Umbral formalism for complex Hermite}
Before entering the main topic of this section we prefer to use the argument we have discussed so far to introduce the Hermite-umbral formalism.\\
In the previous sections we defined multi-variable ($> 2$)/one index Hermite, they satisfy the evolutionary PDE

\begin{equation}\label{evol_pde}
\begin{split}
& \partial_{x_1}H_n^{(m)}(x_1,x_2,...,x_m)= \sum_{s=2}^{m}\partial_{x_s}^sH_n^{(m)}(x_1,x_2,...,x_m)\\
& H_n^{(m)}(x_1,0,...,0)=x^n
\end{split}
\end{equation}
  
\noindent and thus they can be expressed in terms of the operational identity

\begin{equation}\label{Her_pol_53}
H_{n}^{(m)}(x_1,x_2,...,x_m)=e^{\sum_{s=2}^m\partial_{x_s}^s}x^n
\end{equation}
 
\noindent It is evident that this property too can be exploited to treat families of Gaussian integrals involving HKdF and more advanced forms.\\
As starting example of this section, we consider the integral

\begin{equation}\label{Gauss_Int_Her_pol_1}
I_n(\delta,a,b)= \int_{-\infty}^{+\infty}H_n(x+\delta,y)e^{-ax^2-by^2}dxdy
\end{equation}
 
\noindent and use the operational notation to write it as

\begin{equation}\label{Gauss_Int_Her_pol_1_op_not}
\begin{split}
& I_n(\delta,a,b)= \int_{-\infty}^{+\infty}e^{-ax^2}\Phi_n(x,\delta)dx\\
& \Phi_n(x,\delta)= \int_{-\infty}^{+\infty}e^{-by^2-y\partial_x^2}dy(x+\delta)^n
\end{split}
\end{equation}
 
\noindent The function $\Phi_n(x,\delta)$  is easily derived as indicated below

\begin{equation}\label{Gauss_Int_Her_pol_1_op_not_1}
\begin{split}
& \Phi_n(x,\delta)=\sqrt{\frac{\pi}{b}} e^{\frac{1}{4b}\partial_x^4}(x+\delta)^n=H_n^{(4)}\left(x+\delta, \frac{1}{4b}\right)=\sqrt{\frac{\pi}{b}}e^{x\frac{\partial}{\partial \delta}}H_n^{(4)}\left(\delta, \frac{1}{4b}\right)=\\
& =\sqrt{\frac{\pi}{b}}e^{x\frac{\partial}{\partial \delta}+\frac{1}{4b}\frac{\partial^4}{\partial \delta^4}}\delta^n
\end{split}
\end{equation}
 
\noindent Therefore, in conclusion, we obtain

\begin{equation}\label{Gauss_Int_Her_pol_1_solut}
\begin{split}
& I_n(\delta,a,b)=\sqrt{\frac{\pi}{b}}\int_{-\infty}^{+\infty}e^{-ax^2}e^{x\frac{\partial}{\partial \delta}+\frac{1}{4b}\frac{\partial^4}{\partial \delta^4}}dx \delta^n=\\
& =\frac{\pi}{\sqrt{ab}}e^{\frac{1}{4a}\frac{\partial^2}{\partial \delta^2}+\frac{1}{4b}\frac{\partial^4}{\partial \delta^4}}\delta^n=\frac{\pi}{\sqrt{ab}}H_n^{(4)}\left(\delta, \frac{1}{4a},0, \frac{1}{4b}\right)
\end{split}
\end{equation}
 
\noindent A more advanced example is just given by

\begin{equation}\label{Gauss_Int_Her_pol_2}
\begin{split}
& I_n(\delta,a_1,a_2,a_3)=\int_{-\infty}^{+\infty}H_n^{(3)}(x_1+\delta,x_2,x_3)e^{-a_1x_1^2-a_2x_2^2-a_3x_3^2} dx_1 dx_2 dx_3=\\
& =\sqrt{\frac{\pi^3}{a_1a_2a_3}}H_n^{(6)}\left(\delta, \frac{1}{4a_1},0, \frac{1}{4a_2},0,\frac{1}{4a_3}\right)
\end{split}
\end{equation}
 
\noindent amenable for a straightforward generalization to the $n-th$ variable case.\\

\noindent In ref. \cite{dat-ger} it has been shown that the umbral formalism (for an introduction to the umbral indicial technicalities see \cite{lic-dat}) allows to cast the Hermite polynomials in the form of a Newton binomial, namely

\begin{equation}\label{Her_pol_new_bin}
\begin{split}
& H_n(x,y))=(x+_y\hat{h})^n \varphi_0\\
& _y\hat{h}^r \varphi_0 = \frac{r!}{\Gamma \left(\frac{r}{2}+1\right)}y^{\frac{r}{2}}\left |\cos \left(\frac{r\pi}{2} \right)\right |
\end{split}
\end{equation}
   
\noindent An analogous treatment has been exploited in a recent interesting paper (see ref. \cite{raz}) where the third order HKdF have been written in umbral form by the use of an extension of the operator  $_y\hat{h}^r$ written in terms of circular functions.\\
In this article we introduce the umbral operator defined as

\begin{equation}\label{Her_pol_umbr_op}
\begin{split}
& H_n^{(m)}(x,y))=(x+\hat{h}_m\sqrt[m]{y})^n \varphi_0\\
& \hat{h}_m^r \varphi_0 = \frac{r!}{\Gamma \left(\frac{r}{m}+1\right)}\delta_{m\lfloor \frac{r}{m}\rfloor, r}
\end{split}
\end{equation}
   
\noindent and write the $m-th$ order Hermite according to

\begin{equation}\label{Her_pol_umbr_op_1}
\begin{split}
& H_n^{(m)}(x,y))=(x+\sqrt[m]{y}\hat{h}_m)^n \varphi_0=\sum_{r=0}^n\binom{n}{r}x^{n-r}y^{\frac{r}{m}}\hat{h}_m^r\varphi_0 =\\
& = \sum_{r=0}^n\binom{n}{r}x^{n-r}\frac{r!}{\Gamma \left(\frac{r}{m}+1 \right)}y^{\frac{r}{m}} \delta_{m\lfloor \frac{r}{m}\rfloor, r}=\sum_{r=0}^n\binom{n}{mr}x^{n-mr}\frac{(mr)!}{\Gamma (r+1)}y^r
\end{split}
\end{equation}
   
\noindent which, clearly, coincides with the definition we have already given in the previous section.\\
It is furthermore worth noting that $\hat{h}_2^r \varphi_0$ are the well-known Hermite numbers \cite{Weiss} (OEIS -A067994), $\hat{h}_3^r \varphi_0$ (the third order Hermite numbers) are listed in OEIS as sequence A101109, the higher order sequences $\hat{h}_{m>3}^r \varphi_0$  are (apparently) not yet listed in OEIS.\\
It is furthermore evident that the operational definition in terms of exponential operators now reduces to a straightforward variable shift identity

\begin{equation}\label{Her_pol_var_shif_id_1}
H_n^{(m)}(x,y)= e^{\sqrt[m]{y}\hat{h}_m \partial_x}x^n \varphi_0
\end{equation}

\noindent If we accept the umbral definition, the complex Hermite can be written as

\begin{equation}\label{Her_pol_var_shif_id_2}
H_n^{(m)}(x_1+ix_2,y)= (x_1+ix_2+\sqrt[m]{y}\hat{h}_m)^n \varphi_0=e^{\sqrt[m]{y}\hat{h}_m \partial_z}(x_1+ix_2)^n \varphi_0
\end{equation}

\noindent which can simplify the study of the relevant formal properties.\\
Regarding e. g. the derivative of the polynomials \eqref{Her_pol_var_shif_id_2} with respect to the variable $y$, we find

\begin{equation}\label{Her_pol_deriv_1}
\partial_yH_n^{(m)}(z,y)=\frac{n}{m}y^{\frac{1-m}{m}}\hat{h}_m (z+\sqrt[m]{y}\hat{h}_m)^{n-1} \varphi_0
\end{equation}

\noindent On the other side it is evident that, by the use of the definition \eqref{Her_pol_new_bin}, we end up with

\begin{equation}\label{Her_pol_deriv_2}
e^{\sqrt[m]{y}\hat{h}_m\partial_z}z^n \varphi_0= e^{\sqrt[m]{y}\hat{h}_m\partial_z}\varphi_0 z^n = e^{y\partial_z^m}z^n
\end{equation}

\noindent Before closing this article we consider worth to go back to the incomplete Hermite \eqref{Int_as_H4_two_compl_var} \eqref{two_ind_cpl_her_op_form}, which have been exploited to deal with complex variable, defined 

\begin{equation}\label{Her_pol_incompl_1}
\begin{split}
& h_{m,n}(z, \bar{z} | \tau)=m!n!\sum_{r=0}^{min[m,n]}\frac{\tau^rz^{m-r}\bar{z}^{n-r}}{r!(m-r)!(n-r)!} \\
& z=x_1+ix_2,\; \bar{z}=x_1-ix_2
\end{split}
\end{equation}

\noindent The quantity $\tau$  is usually assumed to be 1, here we consider it to be a variable, possibly complex. Albeit a particular case of the two index Hermite, they were studied, in more recently times, by Ito \cite{Ito} and later by other authors \cite{Sha}. 
If note that 

\begin{equation}\label{zzbar_deriv_1}
\begin{split}
& z \bar{z}=x_1^2+x_2^2 \\
& \partial_{z,\bar{z}}=\frac{1}{4}(\partial_{x_1}^2+\partial_{x_2}^2)
\end{split}
\end{equation}

\noindent we argue from the second of eqs. \eqref{two_ind_cpl_her_op_form} that they are solutions of the diffusive equations

\begin{equation}\label{zzbar_deriv_2}
\partial_{\tau}h_{m,n}=\frac{1}{4}(\partial_{x_1}^2+\partial_{x_2}^2)h_{m,n}
\end{equation}

\noindent According to the use of the operational methods discussed so far it is straightforward to conclude that

\begin{equation}\label{zzbar_int_hmn}
\begin{split}
& \int_{-\infty}^{\infty}e^{-ax_1^2-bx_2^2}h_{m,n}(x_1+ix_2, x_1-ix_2 | \tau)dx_1dx_2=\\
& =\frac{\pi}{\sqrt{ab}}H_{m,n}\left(0,\frac{1}{4}\frac{a-b}{ab},0,\frac{1}{4}\frac{a-b}{ab} | \tau +\frac{1}{2}\frac{a+b}{ab}\right)
\end{split}
\end{equation}

\noindent A further element of discussion we like to underscore is that of integrals including products of Hermite polynomials, Gaussian and Bessel functions. We start therefore by considering the example

\begin{equation}\label{int_gaus_bess_1}
_J I_n(x,a)=\int_{-\infty}^{+\infty}e^{-ay^2}J_n(x,y)dy
\end{equation}

\noindent where $J_n(x,y)$ is a two variable Bessel function \cite{dat} specified by the following series expansion and generating function

\begin{equation}\label{two_var_bess_gen_fun_1}
\begin{split}
& J_n(x,y)=\sum_{l=-\infty}^{+\infty}J_{n-2l}(x)J_l(y) \\
& \sum_{n=-\infty}^{+\infty}t^nJ_n(x,y)=e^{\frac{x}{2}\left(t-\frac{1}{t}\right)+\frac{y}{2}\left(t^2-\frac{1}{t^2} \right)}\\
& J_n(z)=\sum_{r=0}^{\infty}(-1)^r\frac{\left(\frac{x}{2}\right)^{n+2r}}{r!(n+r)!} \equiv \mbox{n-th cylindrical Bessel function}
\end{split}
\end{equation}

\noindent The use of the previously outlined methods, yields

\begin{equation}\label{two_var_bess_gen_fun_2}
\begin{split}
& \sum_{n=-\infty}^{+\infty}t^n \; _JI_n(x,y)=e^{\frac{x}{2}\left(t-\frac{1}{t}\right)}\int_{-\infty}^{+\infty}e^{-ay^2}e^{\frac{y}{2}\left(t^2-\frac{1}{t^2} \right)}dy=\\
& =\sqrt{\frac{\pi}{a}}e^{\frac{x}{2}\left(t-\frac{1}{t}\right)+\frac{1}{16a}\left(t^2-\frac{1}{t^2}\right)^2}
\end{split}
\end{equation}

\noindent In order to get a definitive result, further comments are needed. We remind therefore that the generating function  given below

\begin{equation}\label{two_var_bess_gen_fun_3}
\begin{split}
& e^{\frac{x}{2}\left(t-\frac{1}{t}\right)+\frac{y}{2}\left(t^4+\frac{1}{t^4}\right)}=\sum_{n=-\infty}^{+\infty}t^nJI_n^{(4)}(x,y) \\
& JI_n^{(4)}(x,y)=\sum_{l=-\infty}^{+\infty}J_{n-4l}(x)I_l(y)\\
&I_n(y)=\sum_{r=0}^{\infty}\frac{\left(\frac{y}{2}\right)^{n+2r}}{r!(n+r)!} \equiv \mbox{modified Bessel of first kind}
\end{split}
\end{equation}

\noindent Thus finding from eq. \eqref{two_var_bess_gen_fun_2}

\begin{equation}\label{JIn_1}
_J I_n(x,a)=\sqrt{\frac{\pi}{a}}e^{-\frac{1}{8a}}JI_n^{(4)}\left(x,\frac{1}{8a}\right)
\end{equation}

\noindent Within the context of generalized multivariable Bessel an important role is played by the so called Hermite Bessel functions \cite{dat} defined by the operational identity

\begin{equation}\label{her_bes_1}
\begin{split}
&  _H J_n(x,y)=e^{y\partial_x^2}J_n(x)=\\
& =\sum_{r=0}^{\infty}\frac{(-1)^r}{2^{n+2r}}\frac{H_{n+2r}(x,y)}{r!(n+r)!}
\end{split}
\end{equation}

\noindent The last example, we consider worth to discuss in order to stress the flexibility of the method, is

\begin{equation}\label{her_bes_2}
 _{JGH}I_{n,m}(x,z,a)=\int_{-\infty}^{+\infty}e^{-ay^2}H_n(x,y) _HJ_m(z,y)dy
\end{equation}

\noindent and can be explicitly worked out as

\begin{equation}\label{her_bes_3}
\begin{split}
&  _{JGH}I_{n,m}(x,z,a)=\int_{-\infty}^{+\infty}e^{-ay^2}e^{y(\partial_x^2+\partial_z^2)}dy \; x^nJ_m(z)=\\
& = \sqrt{\frac{\pi}{a}}e^{\frac{1}{4a}(\partial_x^2+\partial_z^2)^2}(x^n J_m(z))
\end{split}
\end{equation}

\noindent The final result is provided by the action of the exponential operators on a monomial and on a Bessel function. We note first that

\begin{equation}\label{exp_deriv_2_oper_appl}
\begin{split}
&  e^{\frac{1}{4a}(\partial_x^2+\partial_z^2)^2}(x^nz^m)=H_{n,m}^{(4,4,2)}\left(x,\frac{1}{4a};z,\frac{1}{4a} | \frac{1}{2a}\right) = \\
& =m!n!\sum_{r=0}^{min(\lfloor \frac{m}{2} \rfloor, \lfloor \frac{n}{2} \rfloor)}\left(\frac{1}{2a} \right)^r \; \frac{H_{n-2r}^{(4)}\left(x,\frac{1}{4a}\right)H_{m-2r}^{(4)}\left(z,\frac{1}{4a}\right)}{r!(n-2r)!(m-2r)!}
\end{split}
\end{equation}

\noindent therefore, in conclusion, we end up with

\begin{equation}\label{her_bes_4}
_{JGH}I_{n,m}(x,z,a)=\sqrt{\frac{\pi}{a}}\sum_{s=0}^{\infty}\frac{(-1)^s}{2^{m+2s}s!(n+s)!}H_{n,m+2s}^{(4,4,2)}\left(x, \frac{1}{4a}; z, \frac{1}{4a} | \frac{1}{2a}\right)
\end{equation}

\noindent The last points we have touched on yields an idea of how wide are the implication offered by the methods, we have foreseen so far.\\
In this article we have shown that much progress can be done on the study of different families of Hermite polynomials, by the use of methods developed in the course of the last years. These techniques have benefitted from the contributions grown in different math field of researches and have also been shown to be a powerful tool in applications \cite{bab-dat-lic-sab}\cite{dat-mig}\cite{dat-sol-tor} are within the framework of a reformulation of special functions and polynomials in terms of different types of operational methods, including those of umbral nature.\\
In a forthcoming paper we will apply the methods, we have outlined here, to problems associated with the non-paraxial evolution of elliptical Gaussian beams. An example in this direction is the recently published article in \cite{zhu}, where it has been shown that operators of the type $e^{\alpha \partial_x^2}$ , with $\alpha$  being a complex variable, play a central role in the study of the solution of the Helmholtz equation (see also \cite{dat-sol-tor}).\\

\newenvironment{acknowledgements}%
    {\begin{center}%
    \bfseries Acknowledgements\end{center}}%
    {}
        \begin{acknowledgements}
        This Article is dedicated to the late Dr. Antonio (Tonino) Dipace in recognition of his outstanding mathematical culture and for the delightful discussions we had in forty years of friendship.\\
\; \\
The work of Dr S. Licciardi is supported by the project “Network 4 Energy Sustainable Transition – NEST”, code PE0000021, CUP B73C22001280006, Spoke 7, funded under the National Recovery and Resilience Plan (NRRP), Mission 4, by the European Union – NextGenerationEU.

        \end{acknowledgements}

\vspace{6pt}



\end{document}